\documentclass[aps,prl,twocolumn,superscriptaddress]{revtex4-1}
\usepackage{tabularx}
\usepackage{bm}
\usepackage{euscript}
\usepackage{epsfig,psfrag,subfigure}
\usepackage{graphicx}
\usepackage{color}
\usepackage{amsfonts}
\usepackage{exscale}
\usepackage{amsbsy}
\usepackage{multirow}
\usepackage{hyperref}
\hypersetup{colorlinks=true,linkcolor=black}
\usepackage[]{lineno}
\usepackage{comment}
\usepackage{ulem}
\usepackage[dvipsnames]{xcolor}
\usepackage{amssymb}
\usepackage{amsmath}
\usepackage{natbib}
\usepackage{mathrsfs}
\usepackage{braket}
\usepackage[toc,page,title,titletoc,header]{appendix}

\def\be{\begin{equation}}       \def\ee{\end{equation}}
\def\bea{\begin{eqnarray}}      \def\eea{\end{eqnarray}}

\newcommand{\PreserveBackslash}[1]{\let\temp=\\#1\let\\=\temp}
\newcolumntype{C}[1]{>{\PreserveBackslash\centering}p{#1}}

\begin{document}
\title{Superconductivity in lightly doped Hubbard model on honeycomb lattice}
\author{Cheng Peng}
\affiliation{Stanford Institute for Materials and Energy Sciences, SLAC National Accelerator Laboratory and Stanford University, Menlo Park, California 94025, USA}

\author{D. N. Sheng}
\affiliation{Department of Physics and Astronomy, California State University, Northridge, California 91330, USA}

\author{Hong-Chen Jiang}
\email{hcjiang@stanford.edu}
\affiliation{Stanford Institute for Materials and Energy Sciences, SLAC National Accelerator Laboratory and Stanford University, Menlo Park, California 94025, USA}

\begin{abstract}
We have performed large-scale density-matrix renormalization group studies of the lightly doped Hubbard model on the honeycomb lattice on long three and four-leg cylinders. We find that the ground state of the system upon lightly doping is consistent with that of a superconducting state with coexisting quasi-long-range superconducting and charge density wave orders. Both the superconducting and charge density wave correlations decay as a power law at long distances with corresponding exponents $K_{sc}<2$ and $K_c<2$. On the contrary, the spin-spin and single-particle correlations decay exponentially, although with relatively long correlation lengths. 
\end{abstract}

\maketitle

The Hubbard model is believed as a minimum effective model capturing the low energy physics of the cuprate superconductors \cite{Hubbard2013,Arovas2022,Qin2022}. With strong on-site Coulomb repulsion ($U\gg t$) at half-filling, 
the Hubbard model becomes a Mott insulator where electrons are localized. In cuprates, high-temperature superconductivity can emerge by doping holes into the parent antiferromagnetic Mott insulator \cite{Lee2006,Fradkin2015}. However, it remains an unsettled  issue whether the simplest Hubbard model with nearest-neighbor (NN) electron hopping terms on the square lattice can lead to superconductivity or other additional terms such as next-nearest-neighbor (NNN) electron hopping terms are essential to induce the unconventional superconductivity. While superconductivity appears promising in the hole doped case among other competing states on four-leg square cylinders \cite{PRB1997White,PRB1999White,Jiang2018tJ,Jiang2019hub,Dodaro2017,Chung2020,Jiang2020prb,Jiang2020prr}, systematical density-matrix renormalization group (DMRG) studies on six-leg \cite{Gong2021,Jiang2021PRL} and eight-leg \cite{Jiang2021WT,Jiang2022White} square cylinders suggest the absence of superconductivity in the hole-doped case 
while competing charge density wave correlations dominate in the $t$-$J$ model as the large $U$ limit of the Hubbard model, although strong superconductivity can be realized in the electron-doped cases.\cite{Jiang2021WT,Jiang2021PRL,Jiang2022White,Jiang2023} 
The same issue regarding the fate of superconductivity upon hole doping also  applies to the  Hubbard model on the honeycomb lattice, which has an antiferromagnetic order in the large $U/t$ region at half-filling \cite{zhengchengprb2020}.

The Hubbard model on honeycomb lattice has its own importance as many twisted Moir\'e systems may naturally realize quantum simulators for such a model and its bilayer or multi-component extensions with tunable interactions\cite{Pan2020,Yuan2018}, with the Winger crystal state observed experimentally\cite{moire2021exp,nitin2022}. However, relatively less progress has been made regarding the nature of quantum phases in the honeycomb lattice Hubbard model. Controversies have been raised between different studies regarding whether superconductivity can emerge on the honeycomb lattice when holes are doped into the antiferromagnetic Mott insulating phase. Mean-field and tensor network studies suggest that the antiferromagnetic order near half-filling may coexist with  spin-singlet or/and spin-triplet superconductivity from the perspective of either Hubbard or closely related
$t$-$J$ models\cite{QiY2020,gu2013,xu2022,miao2023}. However, a recent DMRG study reported charge stripe order in the doped Hubbard model on the honeycomb lattice without superconductivity \cite{YangX2021,MingpuPRBstripe}. As a result, it remains unsettled what is the precise nature of the ground state of doped Hubbard model on the honeycomb lattice.

\textit{Principal results -- }%
To address these questions, we have studied the lightly doped single-band Hubbard model on the honeycomb lattice using large-scale DMRG simulations. Our results on long three- and four-leg cylinders on the honeycomb lattice suggest that the ground state of the system is consistent with a superconducting (SC) state with quasi-long-range SC and charge density wave (CDW) correlations, but short-range spin-spin and single-particle correlations. The charge density profile corresponds to a local pattern of partially filled charge stripes with two doped holes in each one-dimensional (1D) CDW unit cell. The spin-singlet SC correlations $\Phi(r)$ are dominant in the pairing channel whose pairing symmetry is consistent with $d$-wave. At long distances, we find that $\Phi(r)\sim r^{-K_{sc}}$ with a Luttinger exponent $K_{sc}<2$. This implies a SC susceptibility that diverges as $\chi_{sc}\sim T^{-(2-K_{sc})}$ as the temperature $T\rightarrow 0$.

\begin{figure}[tb]
\centering
    \includegraphics[width=0.7\linewidth]{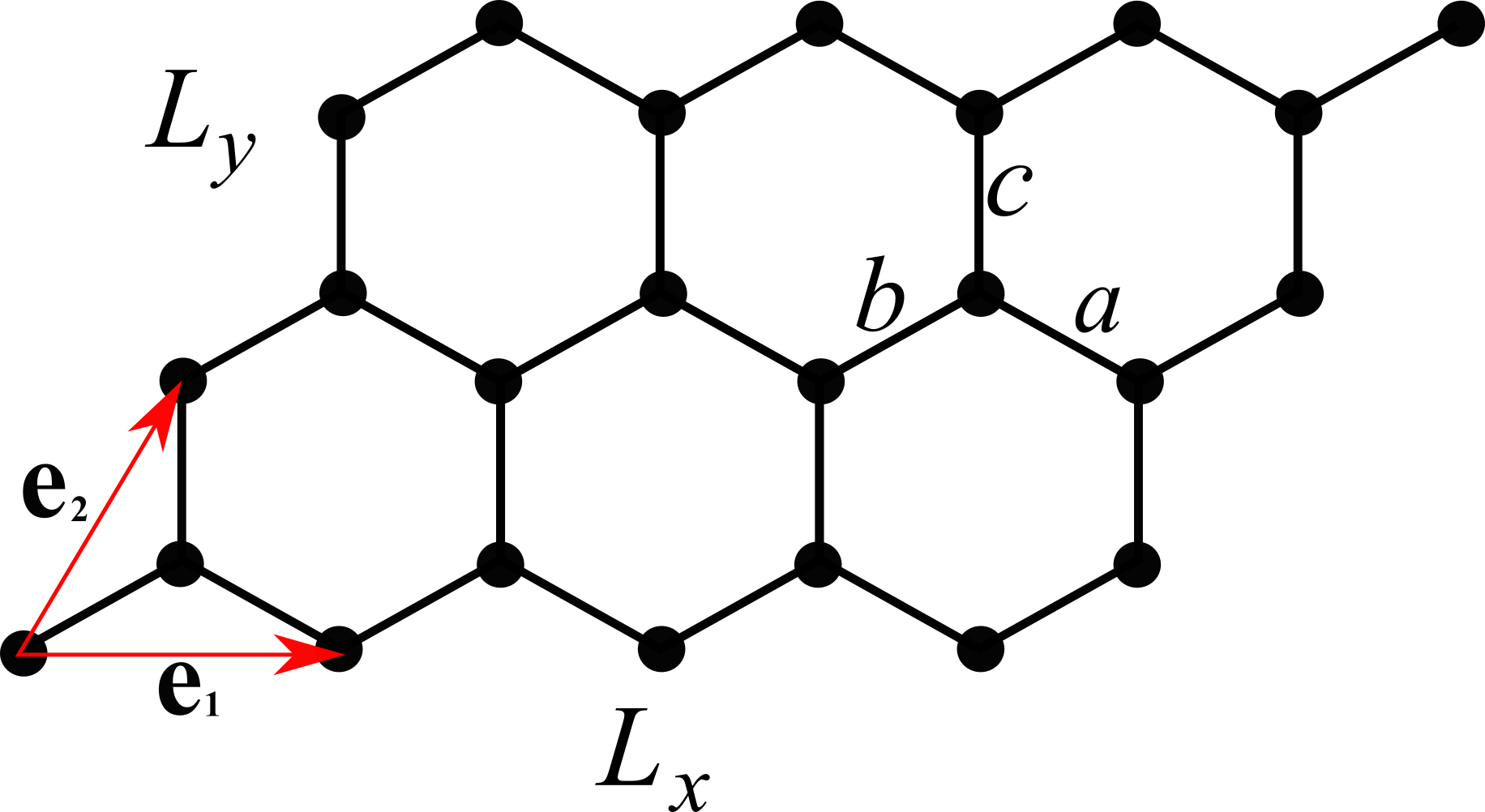}
\caption{(Color online) Schematic three-leg cylinder on the honeycomb lattice. The open (periodic) boundary condition is imposed along the direction specified by the lattice basis vector $\mathbf{e}_1$ ($\mathbf{e}_2$). $L_x$ ($L_y$) is the number of unit cells in the $\mathbf{e}_1$ ($\mathbf{e}_2$) direction. $a$, $b$ and $c$ label the three different bonds.} \label{Fig:lattice}
\end{figure}

\textit{Model and Method -- }%
We use DMRG \cite{White1992} to study the ground state properties of the lightly doped single-band Hubbard model on the honeycomb lattice, whose Hamiltonian is defined as%
\begin{eqnarray} \label{Eq:Ham}
  H= &-& t\sum_{\langle ij\rangle \sigma}\left(\hat{c}^{\dagger}_{i\sigma}\hat{c}_{j\sigma}+h.c.\right) + U\sum_{i}\hat{n}_{i\uparrow}\hat{n}_{i\downarrow} 
\end{eqnarray}
Here, $\hat{c}^{\dagger}_{i\sigma}$ ($\hat{c}_{i\sigma}$) is the electron creation (annihilation) operator with spin-$\sigma$ ($\sigma=\uparrow, \downarrow$) on site $i=(x_i,y_i)$, $\hat{n}_{i\sigma}=\hat{c}^{\dagger}_{i\sigma}\hat{c}_{i\sigma}$ and $\hat{n}_{i}=\sum_{\sigma}\hat{n}_{i,\sigma}$ are the electron number operators. $t$ is the electron hopping amplitude between NN sites $\langle ij\rangle$, and $U$ is the on-site Coulomb repulsion. We take the lattice geometry to be cylindrical, as shown in Fig.\ref{Fig:lattice}, the cylinder has periodic boundary condition along the $\vec{e}_2=(1/2,\sqrt{3}/2)$ direction and open boundary condition along the $\vec{e}_1=(1,0)$ direction. Here, we consider cylinders with circumference $L_y$ and length $L_x$, where $L_y$ and $L_x$ are the number of unit cells along the $\vec{e}_2$ and $\vec{e}_1$ directions, respectively. For three-leg cylinders, i.e., $L_y=3$, the total number of sites is $N=L_x\times L_y\times 2= 2N_u$, where each unit cell has two sites and $N_u$ denotes the number of unit cells. For four-leg cylinders, i.e., $L_y=4$, we have added an additional column on the right open boundary in the practical DMRG calculations to restore the reflection symmetry of the CDW oscillation relatively away from the open boundaries. The corresponding total number of sites on the four-leg cylinder is $N=L_x\times L_y\times 2 + L_y=2N_u+L_y$.

In the present study, we focus primarily on three-leg and four-leg cylinders, i.e., $L_y=3$ and $L_y=4$, with lengths up to $L_x=32$. The doping concentration away from the half-filling is defined as $\delta=N_h/(2N_u)$ where $N_h$ denotes the number of doped holes. For four-leg cylinders, although $N\neq 2N_u$ so that the value of $\delta$ differs slightly from $\tilde{\delta}=N_h/N$ in the vicinity of the open ends, deep in bulk, i.e., relatively away from the boundaries, it is approximate $\delta=\tilde{\delta}$. We focus on lightly doped cases with hole doping concentration $\delta=1/12$ and $1/16$. We set $t=1$ as an energy unit and consider $U=8$ and $U=12$. We perform more than 100 sweeps and keep up to $m=40,000$ in each DMRG block with a typical truncation error $\epsilon\sim 10^{-6}$. Further details of the numerical simulation are provided in the Supplementary Material (SM).

\begin{figure}[htb]
\centering
    \includegraphics[width=1\linewidth]{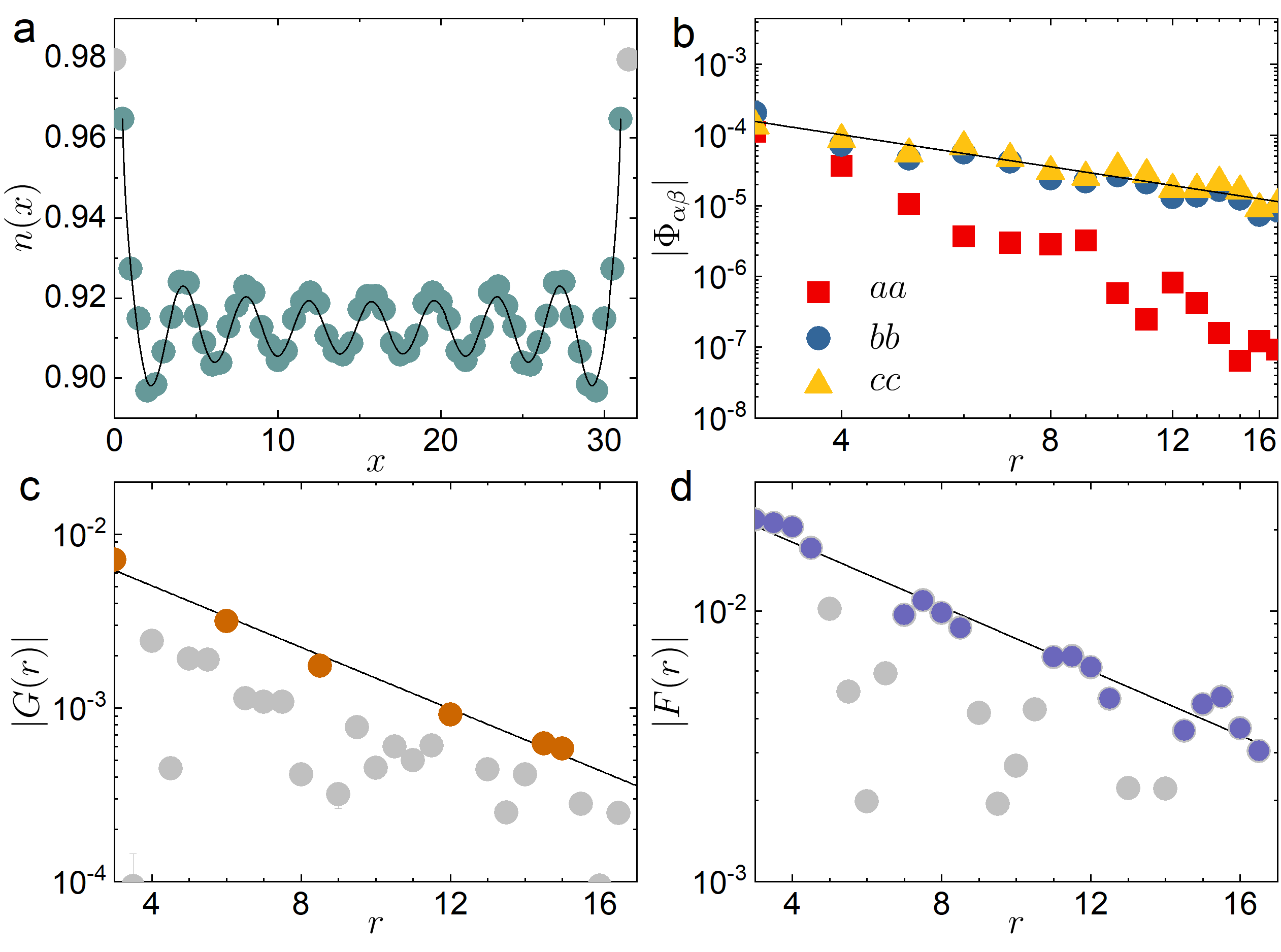}
\caption{(Color online) Correlation function of three-leg cylinder at $1/12$ hole doping with $U = 12$. (a) Charge density profile $n(x)$ where the solid lines denote the fitting using Eq.(\ref{Eq:FriedelOscillation}). Data points in gray are discarded to minimize the boundary effect. (b) Superconducting correlations $|\Phi_{\alpha\beta}(r)|$ with $\alpha\beta=aa$,$bb$ and $cc$. The black line denotes the power-law fitting function $f(r)\sim r^{-K_{sc}}$. (c) Single-particle correlation $|G_{\sigma}(r)|$ and the exponential fitting function $f(r)\sim e^{-r/\xi_G}$ (black line). (d) Spin-spin correlation $|F(r)|$ and the exponential fitting function $f(r)\sim e^{-r/\xi_s}$ (black line). Note that data points far from the envelope or have large error bars are discarded in the fitting process and shown in gray color in (c)-(d).} \label{Fig:3legU12}
\end{figure}

\begin{figure}[htb]
\centering
    \includegraphics[width=1\linewidth]{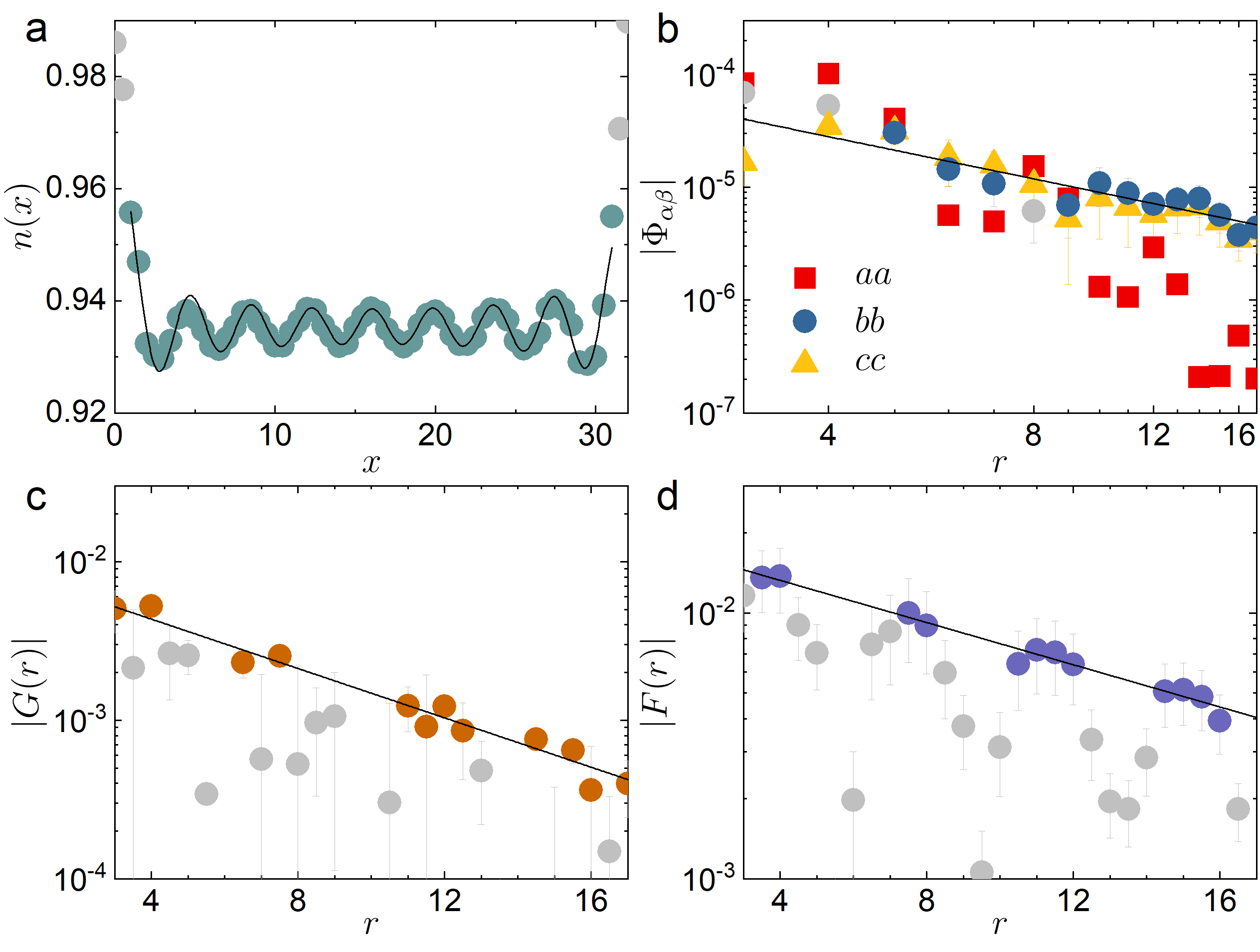}
\caption{(Color online) Correlation function of four-leg cylinder at $1/16$ hole doping with $U = 8$. (a) Charge density profile $n(x)$ where the solid lines denote the fitting using Eq.(\ref{Eq:FriedelOscillation}). Data points in gray are discarded to minimize the boundary effect. (b) Superconducting correlations $|\Phi_{\alpha\beta}(r)|$ with $\alpha\beta=aa$,$bb$ and $cc$. The black line denotes the power-law fitting function $f(r)\sim r^{-K_{sc}}$. (c) Single-particle correlation $|G_{\sigma}(r)|$ and the exponential fitting function $f(r)\sim e^{-r/\xi_G}$ (black line). (d) Spin-spin correlation $|F(r)|$ and the exponential fitting function $f(r)\sim e^{-r/\xi_s}$ (black line). Note that data points far from the envelope or have large error bars are discarded in the fitting process and shown in gray color in (c)-(d).} \label{Fig:4legU8}
\end{figure}

\begin{table*}[tb]
\centering 
\begin{tabular}{C{0.45\linewidth} | C{0.12\linewidth}  C{0.12\linewidth}  C{0.12\linewidth}  C{0.12\linewidth}}  
\hline\hline 
Parameters & $K_c$ & $K_{sc}$ & $\xi_{s}$ & $\xi_{G}$ \\ [.5ex] 
\hline 
$U=12$, $\delta=1/12$, $N=32\times3\times2$, $m$ up to $30000$ & $0.98(8)$ & $1.50(8)$ & $7.3(4)$ & $4.9(3)$ \\
$U=8$, $\delta=1/16$, $N=32\times4\times2+4$, $m$ up to $40000$ & $1.3(2)$ & $1.2(1)$ & $10.9(4)$ & $5.6(3)$ \\
\hline\hline 
\end{tabular}
\caption{The table lists the lattice parameters, Luttinger exponents ($K_c$, $K_{sc}$), and correlation lengths ($\xi_s$, $\xi_G$) in the unit of unit cell. We set t = 1 as an energy unit. Two different on-site repulsions $U$ and hole doping concentrations $\delta$ are considered.}\label{Table:Exponent}
\end{table*}

\textit{Charge density wave order -- }%
To describe the charge density properties of the ground state of the system, we have calculated the charge density profile $n(x,y)=\langle \hat{n}(x,y)\rangle$ on-site $i=(x,y)$ and its rung average $n(x)=\sum_{y=1}^{L_y}n(x,y)/L_y$, where $x$ is the rung index of the cylinder in the unit of $\mathbf{e}_{1}$. Note that there are two sites for each unit cell with the cell parameter $\mathbf{e}_1$ and $\mathbf{e}_2$. Our results show that the system forms ``partially filled" charge stripes with two doped holes in each CDW unit cell (to sum the hole density of all legs). Specifically, the wavelength of charge stripes, i.e., the spacing between two adjacent stripes along the $\mathbf{e}_1$ direction is $\lambda_{c}=1/3\delta$, i.e., $\lambda_c=4$ at $\delta=1/12$, in the unit of $\mathbf{e}_1$ on three-leg cylinders as shown in Fig.\ref{Fig:3legU12}a. This corresponds to an ordering wave vector $Q=6\pi \delta$. For four-leg cylinders, as shown in Fig.\ref{Fig:4legU8}a, the charge stripes have a CDW wavelength $\lambda_{c}=1/4\delta$, i.e., $\lambda_c=4$ at $\delta=1/16$, and two doped holes per each 1D CDW unit cell. Such a ``partially-filled" charge stripe is similar (in the unit of lattice site) with that of the lightly doped Hubbard and $t$-$J$ models on four-leg square cylinders \cite{PRB1997White,PRB1999White,Jiang2018tJ,Jiang2019hub,Chung2020,Jiang2020prb,Jiang2020prr,Jiang2021PRL}.
 
At long distances, our results show that the spatial decay of the CDW correlations is dominated by a power-law with a Luttinger exponent $K_c$. Numerically, the exponent $K_c$ can be obtained by fitting the charge density oscillations (Friedel oscillations) induced by open boundaries of the cylinder\cite{White2002,cdwosc2015prb}%
\begin{eqnarray}\label{Eq:FriedelOscillation}
n(x) &\approx& \frac{A\cos(Qx+\phi_{1})}{[L_{\text{eff}} \sin(\pi x/L_{\text{eff}}+\phi_{2})]^{K_c/2}}+n_0.
\end{eqnarray}
Here $A$ is a non-universal amplitude, $\phi_{1}$ and $\phi_{2}$ are the phase shifts, and $n_0$ is the average charge density. Examples of the fitting using Eq.(\ref{Eq:FriedelOscillation}) are shown in Fig.\ref{Fig:3legU12}a for the three-leg cylinder at $\delta=1/12$ hole doping and Fig.\ref{Fig:4legU8}a for the four-leg cylinder at $\delta=1/16$. Note that four data points close to the open boundaries are excluded in the fitting process to minimize the boundary effect. The extracted exponent $K_c\sim 1$ whose precise values are provided in Table \ref{Table:Exponent}. Similarly, the exponent $K_c$ can also be extracted from the charge density-density correlations, which gives qualitatively consistent results. The details are provided in the SM. We compare our results with previous DMRG study \cite{YangX2021}; we observe consistency for the local pattern of charge stripes $n(x)$, but disagreement on its long-distance decaying behavior, which might be attributed to the fact that we have kept a significantly larger number of states in the DMRG calculations and considered noticeably longer cylinders.

\textit{Superconducting correlations -- }%
To test the possibility of superconductivity, we have calculated the equal-time spin-singlet SC correlation, which is defined as
\begin{eqnarray}
\Phi_{\alpha\beta}(r)=\langle\Delta^{\dagger}_{\alpha}(x_0,y)\Delta_{\beta}(x_0+r,y)\rangle. \label{Eq:SC}
\end{eqnarray}
Here $\Delta^{\dagger}_{\alpha}(x,y)=[\hat{c}^{\dagger}_{(x,y),\uparrow}\hat{c}^{\dagger}_{(x,y)+\alpha,\downarrow} - \hat{c}^{\dagger}_{(x,y),\downarrow}\hat{c}^{\dagger}_{(x,y)+\alpha,\uparrow}]/\sqrt{2}$ is the spin-singlet SC pair creation operator on bond $\alpha$, where $\alpha=a,b,c$ denotes the bond type as shown in Fig.\ref{Fig:lattice}. ($x_0,y$) is the reference bond located at the peak position of the charge density distribution $n(x)$ and most close to $x_0\sim \tilde{L}_x/4$ to minimize the boundary effect, and $r$ is the distance between two bonds in the $\mathbf{e}_1$ direction.

Fig.\ref{Fig:3legU12}b and Fig.\ref{Fig:4legU8}b show the SC correlation $\Phi_{\alpha\beta}(r)$ for $L_y=3$ and $L_y=4$ cylinders at $\delta=1/12$ and $1/16$ doping levels, respectively. As shown in both figures, our results suggest that the pairing symmetry of SC correlations is consistent with that of a nematic $d$-wave, which is reminiscent of the plaquette $d$-wave of the lightly doped $t$-$J$ and Hubbard models on four-leg square cylinders\cite{Dodaro2017,Chung2020}, partially due to the lattice rotational symmetry breaking of the cylindrical geometry. For instance, we find that the SC correlations are dominant on $b$ and $c$ bonds but notably weaker on the $a$ bonds, i.e., $|\Phi_{bb}(r)|\sim |\Phi_{cc}(r)| \gg |\Phi_{aa}(r)|$ on both the $L_y=3$ and $L_y=4$ cylinders. Meanwhile, the SC correlations change sign between different bonds, e.g., $\Phi_{bc}(r)<0$.

At long distances, the dominant SC correlation $\Phi(r)$, e.g., $\Phi_{bb}(r)$ and $\Phi_{cc}(r)$, is characterized by a power law with an appropriate Lutinger exponent $K_{sc}$ which is defined as%
\begin{eqnarray}
\Phi(r)\sim r^{-K_{sc}}. \label{Eq:Ksc}
\end{eqnarray}
The extracted exponent $K_{sc}$ by fitting the results in Fig.\ref{Fig:3legU12}b and Fig.\ref{Fig:4legU8}b is provided in Table \ref{Table:Exponent}. A slow decay of the SC correlation with an exponent $K_{sc} < 2$ implies a SC susceptibility that diverges as $\chi\sim T^{-(2-K_{sc})}$ as the temperature $T \rightarrow 0$. This establishes that the lightly doped Hubbard model on both $L_y=3$ and $L_y=4$ cylinders has quasi-long-range SC correlations.
We have also calculated the spin-triplet SC correlations but found that they are much weaker than the spin-singlet SC correlations, as shown in the SM. This suggests that $p$-wave or $p\pm ip$-wave superconductivity is less likely.

\begin{figure}[tb]
\centering
    \includegraphics[width=0.8\linewidth]{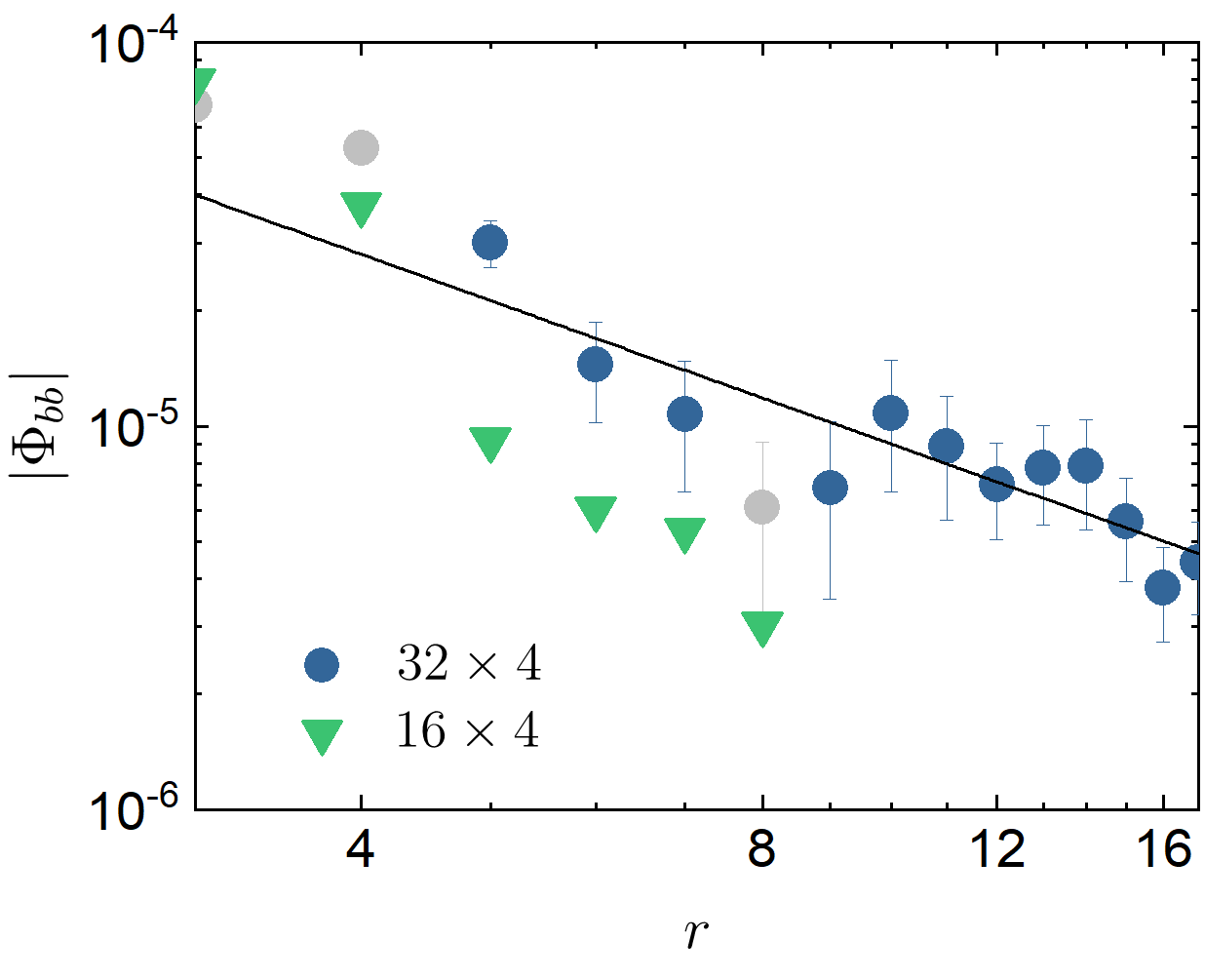}
\caption{(Color online) Superconducting correlations $|\Phi_{bb}(r)|$ for $1/16$ hole doping with $U = 8$. Two different cylinders are considered for comparing the decay behavior.} \label{Fig:comp}
\end{figure}

It is worth mentioning that in order to reliably determine the long-distance decaying behavior of various correlation functions, both long cylinders and a large number of states are required in the DMRG calculations to reduce the finite-size and boundary effects. As an example, we have shown in Fig.\ref{Fig:comp} to compare the dominant SC correlation $\Phi_{bb}(r)$ on four-leg cylinders at $\delta=1/16$ doping concentration by keeping up to $m=40,000$ number of states. It is clear that while the SC correlation decays notably faster on the shorter cylinder due to a stronger boundary effect, it decays much slower on the longer cylinder, which has a smaller boundary effect and is consistent with a power-law decay at long distances.

\textit{Single-particle and spin-spin correlations -- }%
We have also calculated the single-particle correlation function $G_\sigma(r)=\langle c^{\dagger}_{(x_0,y),\sigma} c_{(x_0+r,y)\sigma}\rangle$. Fig.\ref{Fig:3legU12}c and Fig.\ref{Fig:4legU8}c show $G_\sigma(r)$ for $L_y=3$ and $L_y=4$ cylinders at two different doping concentrations, respectively. Our results show that the long-distance behavior of $G_\sigma(r)$ is consistent with an exponential decay $G_\sigma(r)\sim e^{-r/\xi_G}$. The extracted correlation length $\xi_G$ is provided in Table. \ref{Table:Exponent}.

To describe the magnetic properties of the system, we have further calculated the spin-spin correlation function $F(r)=\langle S_{(x_0,y)} S_{(x_0+r,y)}\rangle$. Fig.\ref{Fig:3legU12}d and Fig.\ref{Fig:4legU8}d show $F(r)$ for both $L_y=3$ and $L_y=4$ cylinders at two different doping concentrations, respectively. Similar to $G_\sigma(r)$, we find that $F(r)$ decays exponentially as $F(r)\sim e^{-r/\xi_s}$ at long distances, where the correlation length $\xi_s$ is shown in Table. \ref{Table:Exponent}.

\section{Summary and discussion}%
To summarize, we have studied the ground state properties of the lightly doped Hubbard model on long three- and four-leg cylinders on the honeycomb lattice. Based on the numerical results, we conclude that the ground state of the system is consistent with that of an SC state where both the SC and CDW orders coexist which decay as a power law at long distances with corresponding exponents $K_{sc}<2$ and $K_c<2$. On the contrary, our results suggest that there is a finite gap in both the spin and single-particle sectors which is evidenced by the short-range spin-spin and single-particle Green correlations.

It is worth mentioning that while the pairing symmetry is consistent with that of a $d$-wave, its manifestation on finite honeycomb cylinders is different from that on finite triangular cylinders \cite{Jiang2021Tri}. This is because, on the honeycomb lattice, one component of the SC correlations can be orders of magnitude weaker than other components at long distances, i.e., $\Phi_{aa}(r)\ll \Phi_{bb}(r)\approx \Phi_{cc}(r)$. In other words, it is more similar to the plaquette $d$-wave SC in the lightly hole-doped Hubbard model on four-leg square cylinders with finite negative second-neighbor electron hopping term \cite{Jiang2019hub,Jiang2020prr,Chung2020}, where $\Phi_{xx}(r)\ll \Phi_{yy}(r)$. The fact that the quasi-long-range superconductivity can be realized in the lightly doped Hubbard model on the honeycomb lattice with only the nearest-neighbor electron hopping matrix is notably different from the doped Hubbard model on both the square and triangular lattices \cite{Peng2021,Arovas2022,Qin2022,zhu2022}, where the emergence of quasi-long-range superconductivity requires either a finite next-nearest-neighbor electron hopping term in the uniform Hubbard model \cite{Jiang2019hub,Jiang2020prr,Jiang2021PRL,Jiang2021Tri,Chung2020,Huang2022} or a finite spin gap in the striped Hubbard model \cite{Jiang2022pnas}. 

In the present study, we have focused on the lightly doped Hubbard model with only the nearest-neighbor electron hopping term, it will be interesting to study the higher doping case as well as the effect of longer range electron hopping terms, such as second-neighbor electron hopping term, which has been shown to be essential to enhance the superconductivity on the square lattice \cite{Peng2022}. As the Hubbard model can be naturally realized in many twisted Moir\'e systems as well as their bilayer or multi-component extensions with tunable interactions \cite{Pan2020,Yuan2018,nitin2022}, our results may stimulate efforts to search for unconventional superconductivity in the corresponding field.

\textit{Acknowledgement -- }%
H-C.J. was supported by the Department of Energy (DOE), Office of Sciences, Basic Energy Sciences, Materials Sciences, and Engineering Division, under Contract No. DE-AC02-76SF00515. D.N.S. was supported by DOE Office of Sciences under Grant No. DE-FG02-06ER46305. C.P. acknowledges the support of the U.S. Department of Energy (DOE), Office of Science, Basic Energy Sciences under Contract No. DE-AC02-76SF00515 and Grant No. DE-SC0022216. Part of the computing for this project was performed on the Sherlock cluster.

\appendix 
\renewcommand{\thefigure}{A\arabic{figure}}
\setcounter{figure}{0}
\renewcommand{\theequation}{A\arabic{equation}}
\setcounter{equation}{0}
\setcounter{table}{0}
\renewcommand{\thetable}{A\arabic{table}}
\setcounter{section}{0}

\section{Appendix A: More Results on Three and Four-leg Cylinders}
\begin{figure}[htb]
\centering
    \includegraphics[width=1\linewidth]{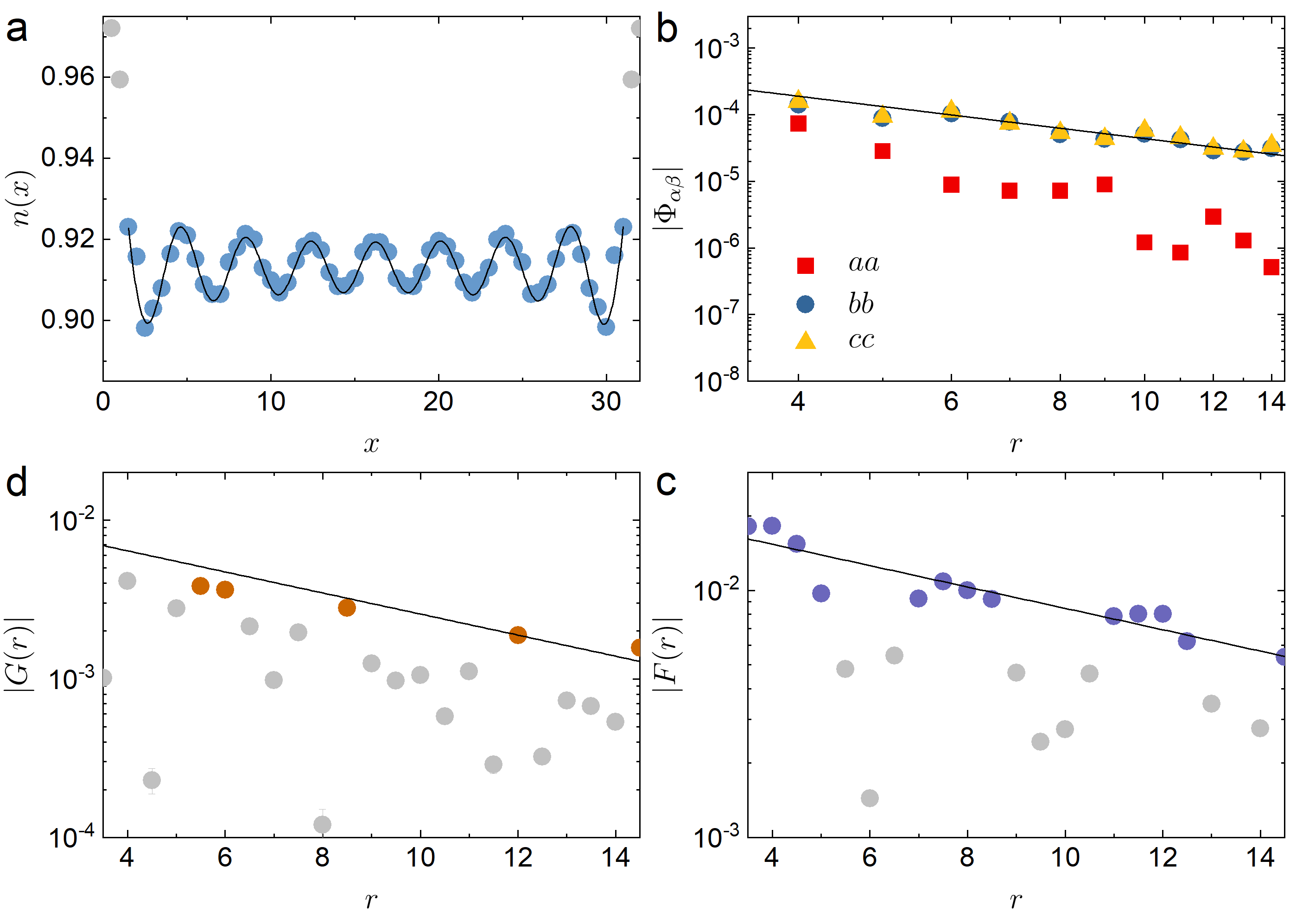}
\caption{(Color online) Correlation function of three-leg cylinder at $1/12$ hole doping with $U/t = 8$. (a) Charge density profile $n(x)$ where the solid lines denote the fitting function. Data points in gray are discarded to minimize the boundary effect. (b) Superconducting correlations $|\Phi_{\alpha\beta}(r)|$ with $\alpha\beta=aa$,$bb$ and $cc$. The black line denotes the power-law fitting function $f(r)\sim r^{-K_{sc}}$. (c) Single-particle correlation $|G_{\sigma}(r)|$ and the exponential fitting function $f(r)\sim e^{-r/\xi_G}$ (black line). (d) Spin-spin correlation $|F(r)|$ and the exponential fitting function $f(r)\sim e^{-r/\xi_s}$ (black line). Note that data points far from the envelope or have large error bars are discarded in the fitting process and shown in gray color in (c)-(d). } \label{Apdx:3legU8}
\end{figure}

\begin{figure}[htb]
\centering
    \includegraphics[width=1\linewidth]{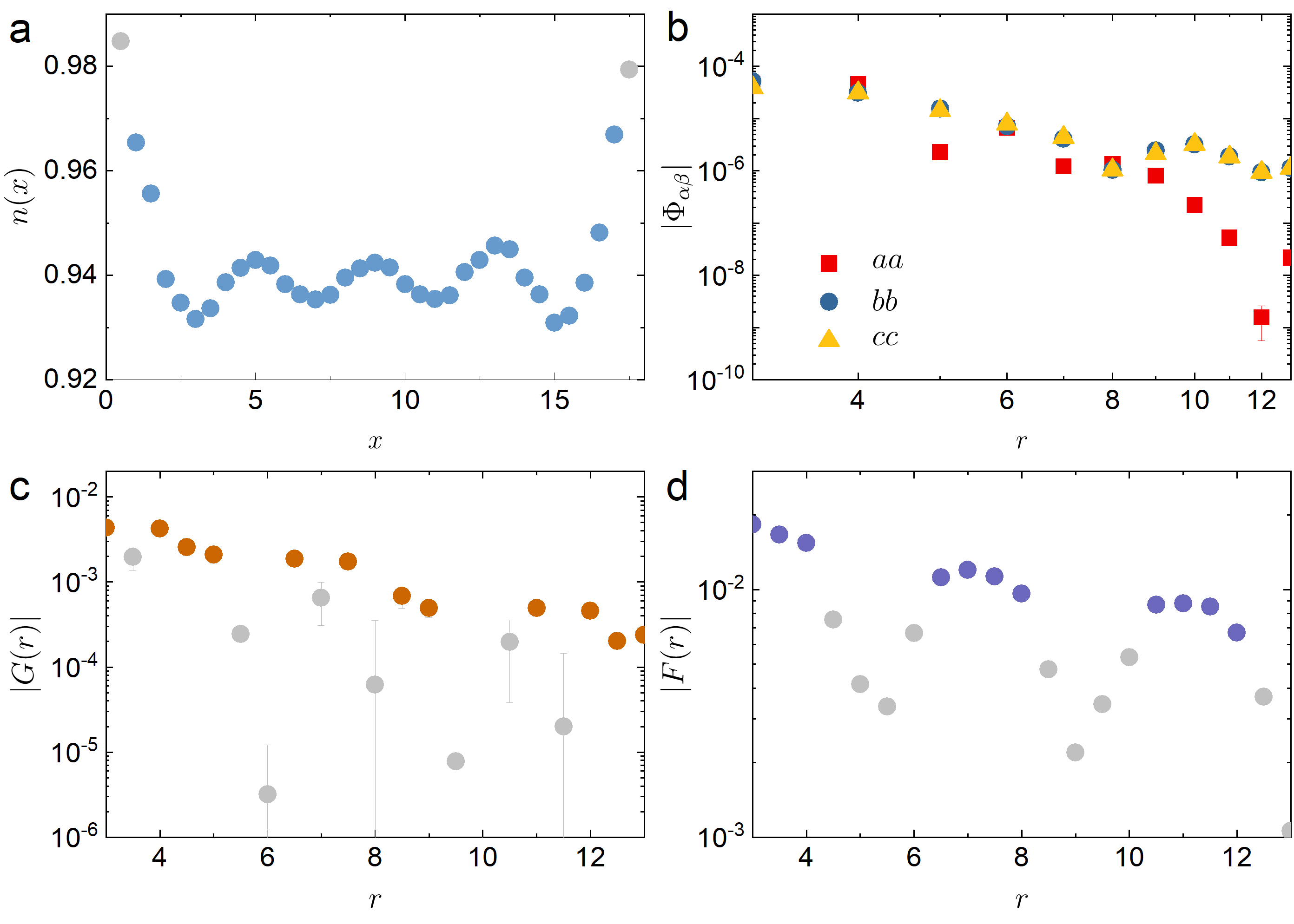}
\caption{(Color online) Correlation function of three-leg cylinder at $1/18$ hole doping with $U/t = 12$. (a) Charge density profile $n(x)$ where the solid lines denote the fitting function. Data points in gray are discarded to minimize the boundary effect. (b) Superconducting correlations $|\Phi_{\alpha\beta}(r)|$ with $\alpha\beta=aa$,$bb$ and $cc$. The black line denotes the power-law fitting function $f(r)\sim r^{-K_{sc}}$. (c) Single-particle correlation $|G_{\sigma}(r)|$ and the exponential fitting function $f(r)\sim e^{-r/\xi_G}$ (black line). (d) Spin-spin correlation $|F(r)|$ and the exponential fitting function $f(r)\sim e^{-r/\xi_s}$ (black line). Note that data points far from the envelope or have large error bars are discarded in the fitting process and shown in gray color in (c)-(d).} \label{Apdx:4legU12}
\end{figure}

We performed DMRG calculations on two more system sizes. Fig.\ref{Apdx:3legU8} displays the results for $U=8$ with $t = 1$ as an energy unit at $1/12$ hole doping concentration on the three-leg cylinder with $L_x=32$. We have kept up to $m=40000$ to fully converge to the ground state. The fitting results are summarized in Table.\ref{Apdx:Exponent}. Compared to the results with different model parameters set up in the main text, the ground state starts to have a very long spin-spin correlation length and a single-particle correlation length. The central charge is $c \sim 3$, which indicates that the ground state deviates from the CDW/superconducting dominant state to become a Luttinger liquid. On the other four-leg system with $U=12$ at $1/18$ hole doping, the results are displayed in Fig.\ref{Apdx:4legU12} and the exponents or correlation lengths are summarized in Table.\ref{Apdx:Exponent}. Since the correlation length is relatively long, roughly half of the system length, that means the middle of the systems can be strongly affected by the boundary behaviors, which we call a boundary effect. Such that on the system smaller than $N=18\times4\times2+4$ may not be adequate to study the single band Hubbard model on the Honeycomb lattice.

\begin{table*}[hb]
\centering 
\begin{tabular}{C{0.45\linewidth} | C{0.12\linewidth}  C{0.12\linewidth}  C{0.12\linewidth}  C{0.12\linewidth}}  
\hline\hline 
Parameters & $K_c$ & $K_{sc}$ & $\xi_{s}$ & $\xi_{G}$ \\ [.5ex] 
\hline 
$U=8$, $\delta=1/12$, $N=32\times3\times2$, $m$ up to $40000$ & $0.98(7)$ & $1.59(7)$ & $13.3(1)$ & $7.5(9)$ \\
$U=12$, $\delta=1/18$, $N=18\times4\times2+4$, $m$ up to $20000$ & $1.3(4)$ & $2.3(2)$ & $11.2(9)$ & $3.2(2)$ \\
\hline\hline 
\end{tabular}
\caption{The table lists the lattice parameters, Luttinger exponents ($K_c$, $K_{sc}$), and correlation lengths ($\xi_G$, $\xi_s$) in the unit of the unit cell at different on-site repulsion $U$, with $t = 1$ as an energy unit, hole doping concentrations $\delta$, and the lattice size $N$.}\label{Apdx:Exponent}
\end{table*}

\section{Appendix B: Density-density correlation function}
The density fluctuation correlation function, defined as $D(r)=\langle [\hat{n}(x_0)-\langle \hat{n}(x_0)\rangle][\hat{n}(x_0+r)-\langle \hat{n}(x_0+r)\rangle]\rangle$, decays in power-law with the same Luttinger exponent $K_c$ as the power. The fitted $K_c$ is $1.46(4)$ for $U=12$ at $\delta=1/12$ hole doping, and $1.57(6)$ for $U=8$ at $\delta=1/16$ hole doping. Note that the $K_c$ given by the density-density correlation function is bigger than that fitted from the CDW; however, it still satisfies $K_c < 2$.

\begin{figure}[htb]
\centering
    \includegraphics[width=1\linewidth]{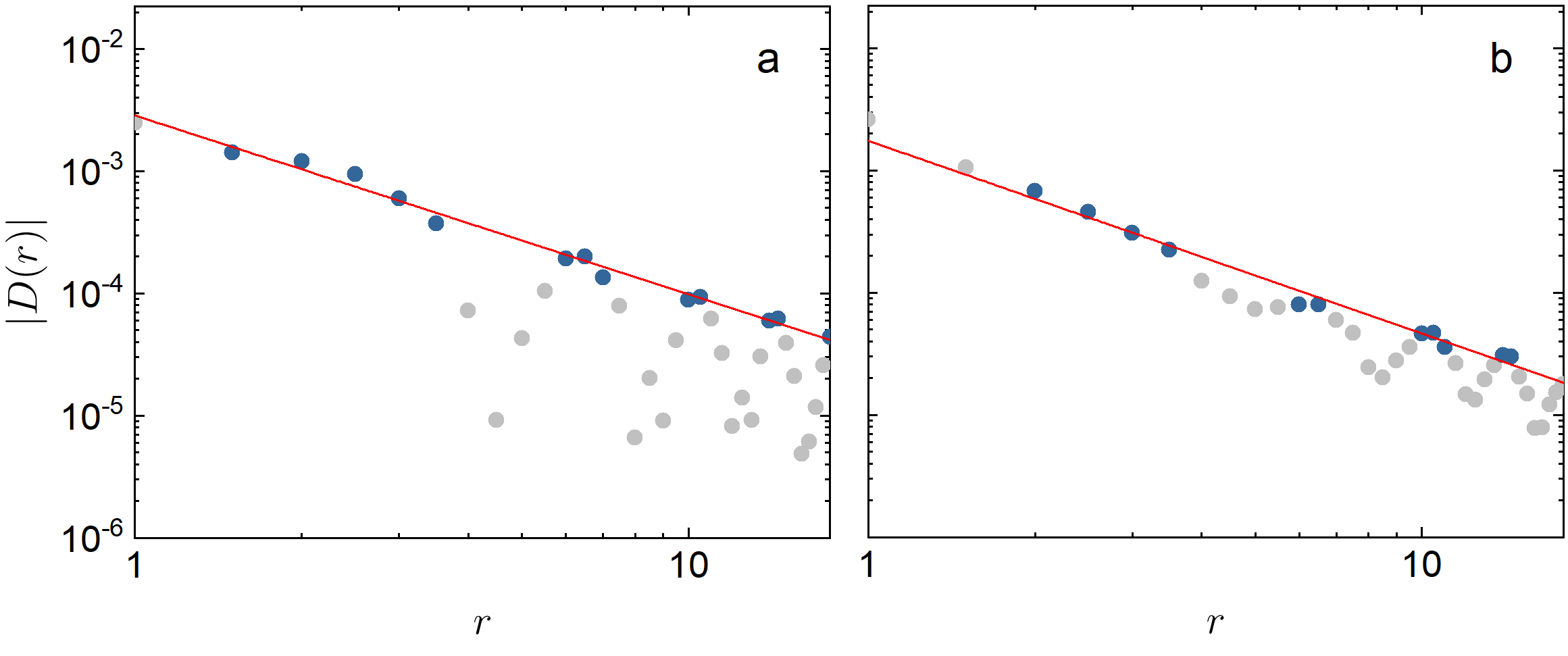}
\caption{(Color online) The density-density correlation function for (a) $U=12$ at $\delta=1/12$ hole doping and (b) $U=8$ at $\delta=1/16$ hole doping. The gray data points are eliminated from fittings.} \label{Apdx:density-density}
\end{figure}

\section{Appendix C: Spin-triplet superconducting correlation function}
We measure the spin-triplet superconducting correlation function defined as
\begin{eqnarray}
\Phi^{triplet}_{\alpha\beta}(r)=\langle\Delta^{\dagger}_{triplet,\alpha}(x_0,y)\Delta_{triplet,\beta}(x_0+r,y)\rangle,
\label{Eq:TC}
\end{eqnarray}
where $\Delta^{\dagger}_{triplet,\alpha}(x,y)=\frac{1}{\sqrt{2}}[c^{\dagger}_{(x,y),\uparrow}c^{\dagger}_{(x,y)+\alpha,\downarrow}+c^{\dagger}_{(x,y),\downarrow}c^{\dagger}_{(x,y)+\alpha,\uparrow}]$. $\alpha=a,b,c$ labels the bond orientations as defined in the main text. Compared with the even-parity SC correlation $\Phi(r)$, $\Phi^{triplet}_{\alpha\beta}(r)$ shown in Supplementary Figure \ref{Apdx:Oddpair} are much weaker which decay faster for the three-leg cylinder at $\delta=1/12$ doping and four-leg cylinder $\delta=1/16$ doping.

\begin{figure}[htb]
\centering
    \includegraphics[width=1\linewidth]{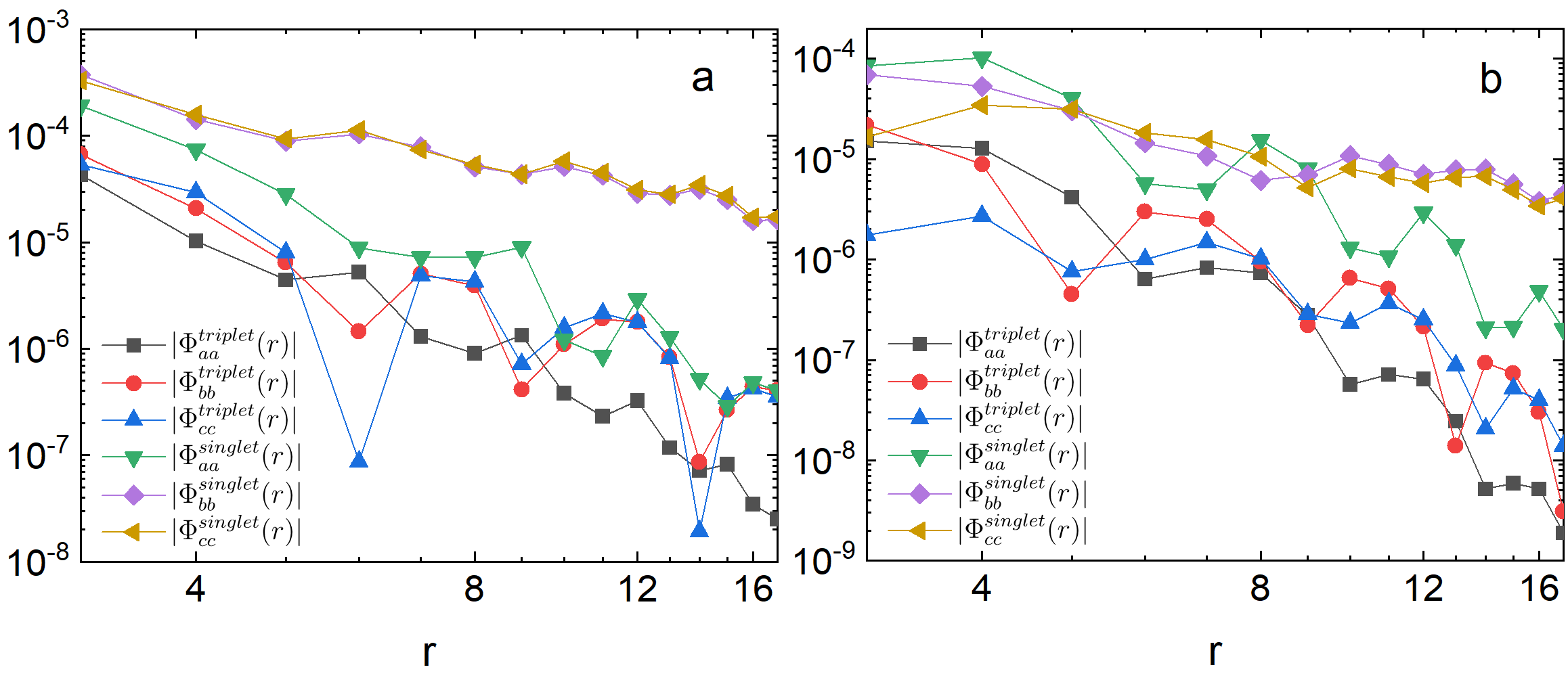}
\caption{(Color online) Various types of SC correlations on (a) three-leg cylinder with $\delta=1/12$ and (b) four-leg cylinder with $\delta=1/16$. We set $U=8$ for two cases.} \label{Apdx:Oddpair}
\end{figure}


\begin{thebibliography}{38}%
\makeatletter
\providecommand \@ifxundefined [1]{%
 \@ifx{#1\undefined}
}%
\providecommand \@ifnum [1]{%
 \ifnum #1\expandafter \@firstoftwo
 \else \expandafter \@secondoftwo
 \fi
}%
\providecommand \@ifx [1]{%
 \ifx #1\expandafter \@firstoftwo
 \else \expandafter \@secondoftwo
 \fi
}%
\providecommand \natexlab [1]{#1}%
\providecommand \enquote  [1]{``#1''}%
\providecommand \bibnamefont  [1]{#1}%
\providecommand \bibfnamefont [1]{#1}%
\providecommand \citenamefont [1]{#1}%
\providecommand \href@noop [0]{\@secondoftwo}%
\providecommand \href [0]{\begingroup \@sanitize@url \@href}%
\providecommand \@href[1]{\@@startlink{#1}\@@href}%
\providecommand \@@href[1]{\endgroup#1\@@endlink}%
\providecommand \@sanitize@url [0]{\catcode `\\12\catcode `\$12\catcode
  `\&12\catcode `\#12\catcode `\^12\catcode `\_12\catcode `\%12\relax}%
\providecommand \@@startlink[1]{}%
\providecommand \@@endlink[0]{}%
\providecommand \url  [0]{\begingroup\@sanitize@url \@url }%
\providecommand \@url [1]{\endgroup\@href {#1}{\urlprefix }}%
\providecommand \urlprefix  [0]{URL }%
\providecommand \Eprint [0]{\href }%
\providecommand \doibase [0]{http://dx.doi.org/}%
\providecommand \selectlanguage [0]{\@gobble}%
\providecommand \bibinfo  [0]{\@secondoftwo}%
\providecommand \bibfield  [0]{\@secondoftwo}%
\providecommand \translation [1]{[#1]}%
\providecommand \BibitemOpen [0]{}%
\providecommand \bibitemStop [0]{}%
\providecommand \bibitemNoStop [0]{.\EOS\space}%
\providecommand \EOS [0]{\spacefactor3000\relax}%
\providecommand \BibitemShut  [1]{\csname bibitem#1\endcsname}%
\let\auto@bib@innerbib\@empty
\bibitem [{\citenamefont {Editorial}(2013)}]{Hubbard2013}%
  \BibitemOpen
  \bibfield  {author} {\bibinfo {author} {\bibnamefont {Editorial}},\ }\href
  {\doibase https://doi-org.stanford.idm.oclc.org/10.1038/nphys2759} {\bibfield
   {journal} {\bibinfo  {journal} {Nat. Phys.}\ }\textbf {\bibinfo {volume}
  {9}},\ \bibinfo {pages} {523} (\bibinfo {year} {2013})}\BibitemShut {NoStop}%
\bibitem [{\citenamefont {Arovas}\ \emph {et~al.}(2022)\citenamefont {Arovas},
  \citenamefont {Berg}, \citenamefont {Kivelson},\ and\ \citenamefont
  {Raghu}}]{Arovas2022}%
  \BibitemOpen
  \bibfield  {author} {\bibinfo {author} {\bibfnamefont {D.~P.}\ \bibnamefont
  {Arovas}}, \bibinfo {author} {\bibfnamefont {E.}~\bibnamefont {Berg}},
  \bibinfo {author} {\bibfnamefont {S.~A.}\ \bibnamefont {Kivelson}}, \ and\
  \bibinfo {author} {\bibfnamefont {S.}~\bibnamefont {Raghu}},\ }\href
  {\doibase 10.1146/annurev-conmatphys-031620-102024} {\bibfield  {journal}
  {\bibinfo  {journal} {Annual Review of Condensed Matter Physics}\ }\textbf
  {\bibinfo {volume} {13}},\ \bibinfo {pages} {239} (\bibinfo {year} {2022})},\
  \Eprint
  {http://arxiv.org/abs/https://doi.org/10.1146/annurev-conmatphys-031620-102024}
  {https://doi.org/10.1146/annurev-conmatphys-031620-102024} \BibitemShut
  {NoStop}%
\bibitem [{\citenamefont {Qin}\ \emph {et~al.}(2022)\citenamefont {Qin},
  \citenamefont {Sch\"{a}fer}, \citenamefont {Andergassen}, \citenamefont
  {Corboz},\ and\ \citenamefont {Gull}}]{Qin2022}%
  \BibitemOpen
  \bibfield  {author} {\bibinfo {author} {\bibfnamefont {M.}~\bibnamefont
  {Qin}}, \bibinfo {author} {\bibfnamefont {T.}~\bibnamefont {Sch\"{a}fer}},
  \bibinfo {author} {\bibfnamefont {S.}~\bibnamefont {Andergassen}}, \bibinfo
  {author} {\bibfnamefont {P.}~\bibnamefont {Corboz}}, \ and\ \bibinfo {author}
  {\bibfnamefont {E.}~\bibnamefont {Gull}},\ }\href {\doibase
  10.1146/annurev-conmatphys-090921-033948} {\bibfield  {journal} {\bibinfo
  {journal} {Annual Review of Condensed Matter Physics}\ }\textbf {\bibinfo
  {volume} {13}},\ \bibinfo {pages} {275} (\bibinfo {year} {2022})},\ \Eprint
  {http://arxiv.org/abs/https://doi.org/10.1146/annurev-conmatphys-090921-033948}
  {https://doi.org/10.1146/annurev-conmatphys-090921-033948} \BibitemShut
  {NoStop}%
\bibitem [{\citenamefont {Lee}\ \emph {et~al.}(2006)\citenamefont {Lee},
  \citenamefont {Nagaosa},\ and\ \citenamefont {Wen}}]{Lee2006}%
  \BibitemOpen
  \bibfield  {author} {\bibinfo {author} {\bibfnamefont {P.~A.}\ \bibnamefont
  {Lee}}, \bibinfo {author} {\bibfnamefont {N.}~\bibnamefont {Nagaosa}}, \ and\
  \bibinfo {author} {\bibfnamefont {X.-G.}\ \bibnamefont {Wen}},\ }\href
  {\doibase 10.1103/RevModPhys.78.17} {\bibfield  {journal} {\bibinfo
  {journal} {Rev. Mod. Phys.}\ }\textbf {\bibinfo {volume} {78}},\ \bibinfo
  {pages} {17} (\bibinfo {year} {2006})}\BibitemShut {NoStop}%
\bibitem [{\citenamefont {Fradkin}\ \emph {et~al.}(2015)\citenamefont
  {Fradkin}, \citenamefont {Kivelson},\ and\ \citenamefont
  {Tranquada}}]{Fradkin2015}%
  \BibitemOpen
  \bibfield  {author} {\bibinfo {author} {\bibfnamefont {E.}~\bibnamefont
  {Fradkin}}, \bibinfo {author} {\bibfnamefont {S.~A.}\ \bibnamefont
  {Kivelson}}, \ and\ \bibinfo {author} {\bibfnamefont {J.~M.}\ \bibnamefont
  {Tranquada}},\ }\href {\doibase 10.1103/RevModPhys.87.457} {\bibfield
  {journal} {\bibinfo  {journal} {Rev. Mod. Phys.}\ }\textbf {\bibinfo {volume}
  {87}},\ \bibinfo {pages} {457} (\bibinfo {year} {2015})}\BibitemShut
  {NoStop}%
\bibitem [{\citenamefont {White}\ and\ \citenamefont
  {Scalapino}(1997)}]{PRB1997White}%
  \BibitemOpen
  \bibfield  {author} {\bibinfo {author} {\bibfnamefont {S.~R.}\ \bibnamefont
  {White}}\ and\ \bibinfo {author} {\bibfnamefont {D.~J.}\ \bibnamefont
  {Scalapino}},\ }\href {\doibase 10.1103/PhysRevB.55.R14701} {\bibfield
  {journal} {\bibinfo  {journal} {Phys. Rev. B}\ }\textbf {\bibinfo {volume}
  {55}},\ \bibinfo {pages} {R14701} (\bibinfo {year} {1997})}\BibitemShut
  {NoStop}%
\bibitem [{\citenamefont {White}\ and\ \citenamefont
  {Scalapino}(1999)}]{PRB1999White}%
  \BibitemOpen
  \bibfield  {author} {\bibinfo {author} {\bibfnamefont {S.~R.}\ \bibnamefont
  {White}}\ and\ \bibinfo {author} {\bibfnamefont {D.~J.}\ \bibnamefont
  {Scalapino}},\ }\href {\doibase 10.1103/PhysRevB.60.R753} {\bibfield
  {journal} {\bibinfo  {journal} {Phys. Rev. B}\ }\textbf {\bibinfo {volume}
  {60}},\ \bibinfo {pages} {R753} (\bibinfo {year} {1999})}\BibitemShut
  {NoStop}%
\bibitem [{\citenamefont {Jiang}\ \emph {et~al.}(2018)\citenamefont {Jiang},
  \citenamefont {Weng},\ and\ \citenamefont {Kivelson}}]{Jiang2018tJ}%
  \BibitemOpen
  \bibfield  {author} {\bibinfo {author} {\bibfnamefont {H.-C.}\ \bibnamefont
  {Jiang}}, \bibinfo {author} {\bibfnamefont {Z.-Y.}\ \bibnamefont {Weng}}, \
  and\ \bibinfo {author} {\bibfnamefont {S.~A.}\ \bibnamefont {Kivelson}},\
  }\href {\doibase 10.1103/PhysRevB.98.140505} {\bibfield  {journal} {\bibinfo
  {journal} {Phys. Rev. B}\ }\textbf {\bibinfo {volume} {98}},\ \bibinfo
  {pages} {140505} (\bibinfo {year} {2018})}\BibitemShut {NoStop}%
\bibitem [{\citenamefont {Jiang}\ and\ \citenamefont
  {Devereaux}(2019)}]{Jiang2019hub}%
  \BibitemOpen
  \bibfield  {author} {\bibinfo {author} {\bibfnamefont {H.-C.}\ \bibnamefont
  {Jiang}}\ and\ \bibinfo {author} {\bibfnamefont {T.~P.}\ \bibnamefont
  {Devereaux}},\ }\href {\doibase 10.1126/science.aal5304} {\bibfield
  {journal} {\bibinfo  {journal} {Science}\ }\textbf {\bibinfo {volume}
  {365}},\ \bibinfo {pages} {1424} (\bibinfo {year} {2019})}\BibitemShut
  {NoStop}%
\bibitem [{\citenamefont {Dodaro}\ \emph {et~al.}(2017)\citenamefont {Dodaro},
  \citenamefont {Jiang},\ and\ \citenamefont {Kivelson}}]{Dodaro2017}%
  \BibitemOpen
  \bibfield  {author} {\bibinfo {author} {\bibfnamefont {J.~F.}\ \bibnamefont
  {Dodaro}}, \bibinfo {author} {\bibfnamefont {H.-C.}\ \bibnamefont {Jiang}}, \
  and\ \bibinfo {author} {\bibfnamefont {S.~A.}\ \bibnamefont {Kivelson}},\
  }\href {\doibase 10.1103/PhysRevB.95.155116} {\bibfield  {journal} {\bibinfo
  {journal} {Phys. Rev. B}\ }\textbf {\bibinfo {volume} {95}},\ \bibinfo
  {pages} {155116} (\bibinfo {year} {2017})}\BibitemShut {NoStop}%
\bibitem [{\citenamefont {Chung}\ \emph {et~al.}(2020)\citenamefont {Chung},
  \citenamefont {Qin}, \citenamefont {Zhang}, \citenamefont {Schollw\"ock},\
  and\ \citenamefont {White}}]{Chung2020}%
  \BibitemOpen
  \bibfield  {author} {\bibinfo {author} {\bibfnamefont {C.-M.}\ \bibnamefont
  {Chung}}, \bibinfo {author} {\bibfnamefont {M.}~\bibnamefont {Qin}}, \bibinfo
  {author} {\bibfnamefont {S.}~\bibnamefont {Zhang}}, \bibinfo {author}
  {\bibfnamefont {U.}~\bibnamefont {Schollw\"ock}}, \ and\ \bibinfo {author}
  {\bibfnamefont {S.~R.}\ \bibnamefont {White}} (\bibinfo {collaboration} {The
  Simons Collaboration on the Many-Electron Problem}),\ }\href {\doibase
  10.1103/PhysRevB.102.041106} {\bibfield  {journal} {\bibinfo  {journal}
  {Phys. Rev. B}\ }\textbf {\bibinfo {volume} {102}},\ \bibinfo {pages}
  {041106} (\bibinfo {year} {2020})}\BibitemShut {NoStop}%
\bibitem [{\citenamefont {Jiang}\ \emph
  {et~al.}(2020{\natexlab{a}})\citenamefont {Jiang}, \citenamefont {Chen},\
  and\ \citenamefont {Weng}}]{Jiang2020prb}%
  \BibitemOpen
  \bibfield  {author} {\bibinfo {author} {\bibfnamefont {H.-C.}\ \bibnamefont
  {Jiang}}, \bibinfo {author} {\bibfnamefont {S.}~\bibnamefont {Chen}}, \ and\
  \bibinfo {author} {\bibfnamefont {Z.-Y.}\ \bibnamefont {Weng}},\ }\href
  {\doibase 10.1103/PhysRevB.102.104512} {\bibfield  {journal} {\bibinfo
  {journal} {Phys. Rev. B}\ }\textbf {\bibinfo {volume} {102}},\ \bibinfo
  {pages} {104512} (\bibinfo {year} {2020}{\natexlab{a}})}\BibitemShut
  {NoStop}%
\bibitem [{\citenamefont {Jiang}\ \emph
  {et~al.}(2020{\natexlab{b}})\citenamefont {Jiang}, \citenamefont {Zaanen},
  \citenamefont {Devereaux},\ and\ \citenamefont {Jiang}}]{Jiang2020prr}%
  \BibitemOpen
  \bibfield  {author} {\bibinfo {author} {\bibfnamefont {Y.-F.}\ \bibnamefont
  {Jiang}}, \bibinfo {author} {\bibfnamefont {J.}~\bibnamefont {Zaanen}},
  \bibinfo {author} {\bibfnamefont {T.~P.}\ \bibnamefont {Devereaux}}, \ and\
  \bibinfo {author} {\bibfnamefont {H.-C.}\ \bibnamefont {Jiang}},\ }\href
  {\doibase 10.1103/PhysRevResearch.2.033073} {\bibfield  {journal} {\bibinfo
  {journal} {Phys. Rev. Research}\ }\textbf {\bibinfo {volume} {2}},\ \bibinfo
  {pages} {033073} (\bibinfo {year} {2020}{\natexlab{b}})}\BibitemShut
  {NoStop}%
\bibitem [{\citenamefont {Gong}\ \emph {et~al.}(2021)\citenamefont {Gong},
  \citenamefont {Zhu},\ and\ \citenamefont {Sheng}}]{Gong2021}%
  \BibitemOpen
  \bibfield  {author} {\bibinfo {author} {\bibfnamefont {S.}~\bibnamefont
  {Gong}}, \bibinfo {author} {\bibfnamefont {W.}~\bibnamefont {Zhu}}, \ and\
  \bibinfo {author} {\bibfnamefont {D.~N.}\ \bibnamefont {Sheng}},\ }\href
  {\doibase 10.1103/PhysRevLett.127.097003} {\bibfield  {journal} {\bibinfo
  {journal} {Phys. Rev. Lett.}\ }\textbf {\bibinfo {volume} {127}},\ \bibinfo
  {pages} {097003} (\bibinfo {year} {2021})}\BibitemShut {NoStop}%
\bibitem [{\citenamefont {Jiang}\ and\ \citenamefont
  {Kivelson}(2021)}]{Jiang2021PRL}%
  \BibitemOpen
  \bibfield  {author} {\bibinfo {author} {\bibfnamefont {H.-C.}\ \bibnamefont
  {Jiang}}\ and\ \bibinfo {author} {\bibfnamefont {S.~A.}\ \bibnamefont
  {Kivelson}},\ }\href {\doibase 10.1103/PhysRevLett.127.097002} {\bibfield
  {journal} {\bibinfo  {journal} {Phys. Rev. Lett.}\ }\textbf {\bibinfo
  {volume} {127}},\ \bibinfo {pages} {097002} (\bibinfo {year}
  {2021})}\BibitemShut {NoStop}%
\bibitem [{\citenamefont {Jiang}\ \emph {et~al.}(2021)\citenamefont {Jiang},
  \citenamefont {Scalapino},\ and\ \citenamefont {White}}]{Jiang2021WT}%
  \BibitemOpen
  \bibfield  {author} {\bibinfo {author} {\bibfnamefont {S.}~\bibnamefont
  {Jiang}}, \bibinfo {author} {\bibfnamefont {D.~J.}\ \bibnamefont
  {Scalapino}}, \ and\ \bibinfo {author} {\bibfnamefont {S.~R.}\ \bibnamefont
  {White}},\ }\href {\doibase https://doi.org/10.1073/pnas.2109978118}
  {\bibfield  {journal} {\bibinfo  {journal} {Proc. Natl. Acad. Sci. U.S.A.}\
  }\textbf {\bibinfo {volume} {118}},\ \bibinfo {pages} {e2109978118} (\bibinfo
  {year} {2021})}\BibitemShut {NoStop}%
\bibitem [{\citenamefont {Jiang}\ \emph {et~al.}(2022)\citenamefont {Jiang},
  \citenamefont {Scalapino},\ and\ \citenamefont {White}}]{Jiang2022White}%
  \BibitemOpen
  \bibfield  {author} {\bibinfo {author} {\bibfnamefont {S.}~\bibnamefont
  {Jiang}}, \bibinfo {author} {\bibfnamefont {D.~J.}\ \bibnamefont
  {Scalapino}}, \ and\ \bibinfo {author} {\bibfnamefont {S.~R.}\ \bibnamefont
  {White}},\ }\href {\doibase 10.1103/PhysRevB.106.174507} {\bibfield
  {journal} {\bibinfo  {journal} {Phys. Rev. B}\ }\textbf {\bibinfo {volume}
  {106}},\ \bibinfo {pages} {174507} (\bibinfo {year} {2022})}\BibitemShut
  {NoStop}%
\bibitem [{\citenamefont {Jiang}\ \emph {et~al.}(2023)\citenamefont {Jiang},
  \citenamefont {Kivelson},\ and\ \citenamefont {Lee}}]{Jiang2023}%
  \BibitemOpen
  \bibfield  {author} {\bibinfo {author} {\bibfnamefont {H.-C.}\ \bibnamefont
  {Jiang}}, \bibinfo {author} {\bibfnamefont {S.~A.}\ \bibnamefont {Kivelson}},
  \ and\ \bibinfo {author} {\bibfnamefont {D.-H.}\ \bibnamefont {Lee}},\ }\href
  {\doibase 10.48550/arXiv.2302.11633} {\  (\bibinfo {year} {2023}),\
  10.48550/arXiv.2302.11633}\BibitemShut {NoStop}%
\bibitem [{\citenamefont {Gu}\ \emph {et~al.}(2020)\citenamefont {Gu},
  \citenamefont {Jiang},\ and\ \citenamefont {Baskaran}}]{zhengchengprb2020}%
  \BibitemOpen
  \bibfield  {author} {\bibinfo {author} {\bibfnamefont {Z.-C.}\ \bibnamefont
  {Gu}}, \bibinfo {author} {\bibfnamefont {H.-C.}\ \bibnamefont {Jiang}}, \
  and\ \bibinfo {author} {\bibfnamefont {G.}~\bibnamefont {Baskaran}},\ }\href
  {\doibase 10.1103/PhysRevB.101.205147} {\bibfield  {journal} {\bibinfo
  {journal} {Phys. Rev. B}\ }\textbf {\bibinfo {volume} {101}},\ \bibinfo
  {pages} {205147} (\bibinfo {year} {2020})}\BibitemShut {NoStop}%
\bibitem [{\citenamefont {Pan}\ \emph {et~al.}(2020)\citenamefont {Pan},
  \citenamefont {Wu},\ and\ \citenamefont {Das~Sarma}}]{Pan2020}%
  \BibitemOpen
  \bibfield  {author} {\bibinfo {author} {\bibfnamefont {H.}~\bibnamefont
  {Pan}}, \bibinfo {author} {\bibfnamefont {F.}~\bibnamefont {Wu}}, \ and\
  \bibinfo {author} {\bibfnamefont {S.}~\bibnamefont {Das~Sarma}},\ }\href
  {\doibase 10.1103/PhysRevB.102.201104} {\bibfield  {journal} {\bibinfo
  {journal} {Phys. Rev. B}\ }\textbf {\bibinfo {volume} {102}},\ \bibinfo
  {pages} {201104} (\bibinfo {year} {2020})}\BibitemShut {NoStop}%
\bibitem [{\citenamefont {Yuan}\ and\ \citenamefont {Fu}(2018)}]{Yuan2018}%
  \BibitemOpen
  \bibfield  {author} {\bibinfo {author} {\bibfnamefont {N.~F.~Q.}\
  \bibnamefont {Yuan}}\ and\ \bibinfo {author} {\bibfnamefont {L.}~\bibnamefont
  {Fu}},\ }\href {\doibase 10.1103/PhysRevB.98.045103} {\bibfield  {journal}
  {\bibinfo  {journal} {Phys. Rev. B}\ }\textbf {\bibinfo {volume} {98}},\
  \bibinfo {pages} {045103} (\bibinfo {year} {2018})}\BibitemShut {NoStop}%
\bibitem [{\citenamefont {Jin}\ \emph {et~al.}(2021)\citenamefont {Jin},
  \citenamefont {Tao}, \citenamefont {Li}, \citenamefont {Xu}, \citenamefont
  {Tang}, \citenamefont {Zhu}, \citenamefont {Liu}, \citenamefont {Watanabe},
  \citenamefont {Taniguchi}, \citenamefont {Hone}, \citenamefont {Fu},
  \citenamefont {Shan},\ and\ \citenamefont {Mak}}]{moire2021exp}%
  \BibitemOpen
  \bibfield  {author} {\bibinfo {author} {\bibfnamefont {C.}~\bibnamefont
  {Jin}}, \bibinfo {author} {\bibfnamefont {Z.}~\bibnamefont {Tao}}, \bibinfo
  {author} {\bibfnamefont {T.}~\bibnamefont {Li}}, \bibinfo {author}
  {\bibfnamefont {Y.}~\bibnamefont {Xu}}, \bibinfo {author} {\bibfnamefont
  {Y.}~\bibnamefont {Tang}}, \bibinfo {author} {\bibfnamefont {J.}~\bibnamefont
  {Zhu}}, \bibinfo {author} {\bibfnamefont {S.}~\bibnamefont {Liu}}, \bibinfo
  {author} {\bibfnamefont {K.}~\bibnamefont {Watanabe}}, \bibinfo {author}
  {\bibfnamefont {T.}~\bibnamefont {Taniguchi}}, \bibinfo {author}
  {\bibfnamefont {J.~C.}\ \bibnamefont {Hone}}, \bibinfo {author}
  {\bibfnamefont {L.}~\bibnamefont {Fu}}, \bibinfo {author} {\bibfnamefont
  {J.}~\bibnamefont {Shan}}, \ and\ \bibinfo {author} {\bibfnamefont {K.~F.}\
  \bibnamefont {Mak}},\ }\href {\doibase
  https://doi.org/10.1038/s41563-021-00959-8} {\bibfield  {journal} {\bibinfo
  {journal} {Nature Materials}\ }\textbf {\bibinfo {volume} {20}},\ \bibinfo
  {pages} {940–944} (\bibinfo {year} {2021})}\BibitemShut {NoStop}%
\bibitem [{\citenamefont {{Kaushal}}\ \emph {et~al.}(2022)\citenamefont
  {{Kaushal}}, \citenamefont {{Morales-Dur{\'a}n}}, \citenamefont
  {{MacDonald}},\ and\ \citenamefont {{Dagotto}}}]{nitin2022}%
  \BibitemOpen
  \bibfield  {author} {\bibinfo {author} {\bibfnamefont {N.}~\bibnamefont
  {{Kaushal}}}, \bibinfo {author} {\bibfnamefont {N.}~\bibnamefont
  {{Morales-Dur{\'a}n}}}, \bibinfo {author} {\bibfnamefont {A.~H.}\
  \bibnamefont {{MacDonald}}}, \ and\ \bibinfo {author} {\bibfnamefont
  {E.}~\bibnamefont {{Dagotto}}},\ }\href {\doibase 10.1038/s42005-022-01065-0}
  {\bibfield  {journal} {\bibinfo  {journal} {Communications Physics}\ }\textbf
  {\bibinfo {volume} {5}},\ \bibinfo {eid} {289} (\bibinfo {year} {2022})},\
  \Eprint {http://arxiv.org/abs/2206.10024} {arXiv:2206.10024
  [cond-mat.str-el]} \BibitemShut {NoStop}%
\bibitem [{\citenamefont {Qi}\ \emph {et~al.}(2020)\citenamefont {Qi},
  \citenamefont {Fu}, \citenamefont {Sun},\ and\ \citenamefont {Gu}}]{QiY2020}%
  \BibitemOpen
  \bibfield  {author} {\bibinfo {author} {\bibfnamefont {Y.}~\bibnamefont
  {Qi}}, \bibinfo {author} {\bibfnamefont {L.}~\bibnamefont {Fu}}, \bibinfo
  {author} {\bibfnamefont {K.}~\bibnamefont {Sun}}, \ and\ \bibinfo {author}
  {\bibfnamefont {Z.}~\bibnamefont {Gu}},\ }\href {\doibase
  10.1103/PhysRevB.102.245140} {\bibfield  {journal} {\bibinfo  {journal}
  {Phys. Rev. B}\ }\textbf {\bibinfo {volume} {102}},\ \bibinfo {pages}
  {245140} (\bibinfo {year} {2020})}\BibitemShut {NoStop}%
\bibitem [{\citenamefont {Gu}\ \emph {et~al.}(2013)\citenamefont {Gu},
  \citenamefont {Jiang}, \citenamefont {Sheng}, \citenamefont {Yao},
  \citenamefont {Balents},\ and\ \citenamefont {Wen}}]{gu2013}%
  \BibitemOpen
  \bibfield  {author} {\bibinfo {author} {\bibfnamefont {Z.-C.}\ \bibnamefont
  {Gu}}, \bibinfo {author} {\bibfnamefont {H.-C.}\ \bibnamefont {Jiang}},
  \bibinfo {author} {\bibfnamefont {D.~N.}\ \bibnamefont {Sheng}}, \bibinfo
  {author} {\bibfnamefont {H.}~\bibnamefont {Yao}}, \bibinfo {author}
  {\bibfnamefont {L.}~\bibnamefont {Balents}}, \ and\ \bibinfo {author}
  {\bibfnamefont {X.-G.}\ \bibnamefont {Wen}},\ }\href {\doibase
  10.1103/PhysRevB.88.155112} {\bibfield  {journal} {\bibinfo  {journal} {Phys.
  Rev. B}\ }\textbf {\bibinfo {volume} {88}},\ \bibinfo {pages} {155112}
  (\bibinfo {year} {2013})}\BibitemShut {NoStop}%
\bibitem [{\citenamefont {Xu}\ \emph {et~al.}(2022)\citenamefont {Xu},
  \citenamefont {Gu},\ and\ \citenamefont {Yang}}]{xu2022}%
  \BibitemOpen
  \bibfield  {author} {\bibinfo {author} {\bibfnamefont {Z.-T.}\ \bibnamefont
  {Xu}}, \bibinfo {author} {\bibfnamefont {Z.-C.}\ \bibnamefont {Gu}}, \ and\
  \bibinfo {author} {\bibfnamefont {S.}~\bibnamefont {Yang}},\ }\href {\doibase
  10.48550/ARXIV.2208.13681} {\enquote {\bibinfo {title} {Competing orders in
  the honeycomb lattice $t$-$j$ model},}\ } (\bibinfo {year}
  {2022})\BibitemShut {NoStop}%
\bibitem [{\citenamefont {Miao}\ \emph {et~al.}(2023)\citenamefont {Miao},
  \citenamefont {Yue}, \citenamefont {Zhang}, \citenamefont {Chen},\ and\
  \citenamefont {Gu}}]{miao2023}%
  \BibitemOpen
  \bibfield  {author} {\bibinfo {author} {\bibfnamefont {J.-J.}\ \bibnamefont
  {Miao}}, \bibinfo {author} {\bibfnamefont {Z.-Y.}\ \bibnamefont {Yue}},
  \bibinfo {author} {\bibfnamefont {H.}~\bibnamefont {Zhang}}, \bibinfo
  {author} {\bibfnamefont {W.-Q.}\ \bibnamefont {Chen}}, \ and\ \bibinfo
  {author} {\bibfnamefont {Z.-C.}\ \bibnamefont {Gu}},\ }\href {\doibase
  10.48550/ARXIV.2301.02274} {\enquote {\bibinfo {title} {Spin-charge
  separation and unconventional superconductivity in \textit{t}-\textit{J}
  model on honeycomb lattice},}\ } (\bibinfo {year} {2023})\BibitemShut
  {NoStop}%
\bibitem [{\citenamefont {Yang}\ \emph {et~al.}(2021)\citenamefont {Yang},
  \citenamefont {Zheng},\ and\ \citenamefont {Qin}}]{YangX2021}%
  \BibitemOpen
  \bibfield  {author} {\bibinfo {author} {\bibfnamefont {X.}~\bibnamefont
  {Yang}}, \bibinfo {author} {\bibfnamefont {H.}~\bibnamefont {Zheng}}, \ and\
  \bibinfo {author} {\bibfnamefont {M.}~\bibnamefont {Qin}},\ }\href {\doibase
  10.1103/PhysRevB.103.155110} {\bibfield  {journal} {\bibinfo  {journal}
  {Phys. Rev. B}\ }\textbf {\bibinfo {volume} {103}},\ \bibinfo {pages}
  {155110} (\bibinfo {year} {2021})}\BibitemShut {NoStop}%
\bibitem [{\citenamefont {Qin}(2022)}]{MingpuPRBstripe}%
  \BibitemOpen
  \bibfield  {author} {\bibinfo {author} {\bibfnamefont {M.}~\bibnamefont
  {Qin}},\ }\href {\doibase 10.1103/PhysRevB.105.035111} {\bibfield  {journal}
  {\bibinfo  {journal} {Phys. Rev. B}\ }\textbf {\bibinfo {volume} {105}},\
  \bibinfo {pages} {035111} (\bibinfo {year} {2022})}\BibitemShut {NoStop}%
\bibitem [{\citenamefont {White}(1992)}]{White1992}%
  \BibitemOpen
  \bibfield  {author} {\bibinfo {author} {\bibfnamefont {S.~R.}\ \bibnamefont
  {White}},\ }\href@noop {} {\bibfield  {journal} {\bibinfo  {journal} {Phys.
  Rev. Lett.}\ }\textbf {\bibinfo {volume} {69}},\ \bibinfo {pages} {2863}
  (\bibinfo {year} {1992})}\BibitemShut {NoStop}%
\bibitem [{\citenamefont {White}\ \emph {et~al.}(2002)\citenamefont {White},
  \citenamefont {Affleck},\ and\ \citenamefont {Scalapino}}]{White2002}%
  \BibitemOpen
  \bibfield  {author} {\bibinfo {author} {\bibfnamefont {S.~R.}\ \bibnamefont
  {White}}, \bibinfo {author} {\bibfnamefont {I.}~\bibnamefont {Affleck}}, \
  and\ \bibinfo {author} {\bibfnamefont {D.~J.}\ \bibnamefont {Scalapino}},\
  }\href {\doibase 10.1103/PhysRevB.65.165122} {\bibfield  {journal} {\bibinfo
  {journal} {Phys. Rev. B}\ }\textbf {\bibinfo {volume} {65}},\ \bibinfo
  {pages} {165122} (\bibinfo {year} {2002})}\BibitemShut {NoStop}%
\bibitem [{\citenamefont {Dolfi}\ \emph {et~al.}(2015)\citenamefont {Dolfi},
  \citenamefont {Bauer}, \citenamefont {Keller},\ and\ \citenamefont
  {Troyer}}]{cdwosc2015prb}%
  \BibitemOpen
  \bibfield  {author} {\bibinfo {author} {\bibfnamefont {M.}~\bibnamefont
  {Dolfi}}, \bibinfo {author} {\bibfnamefont {B.}~\bibnamefont {Bauer}},
  \bibinfo {author} {\bibfnamefont {S.}~\bibnamefont {Keller}}, \ and\ \bibinfo
  {author} {\bibfnamefont {M.}~\bibnamefont {Troyer}},\ }\href {\doibase
  10.1103/PhysRevB.92.195139} {\bibfield  {journal} {\bibinfo  {journal} {Phys.
  Rev. B}\ }\textbf {\bibinfo {volume} {92}},\ \bibinfo {pages} {195139}
  (\bibinfo {year} {2015})}\BibitemShut {NoStop}%
\bibitem [{\citenamefont {Jiang}(2021)}]{Jiang2021Tri}%
  \BibitemOpen
  \bibfield  {author} {\bibinfo {author} {\bibfnamefont {H.-C.}\ \bibnamefont
  {Jiang}},\ }\href@noop {} {\bibfield  {journal} {\bibinfo  {journal} {npj
  Quantum Mater.}\ }\textbf {\bibinfo {volume} {6}},\ \bibinfo {pages} {71}
  (\bibinfo {year} {2021})}\BibitemShut {NoStop}%
\bibitem [{\citenamefont {Peng}\ \emph {et~al.}(2021)\citenamefont {Peng},
  \citenamefont {Jiang}, \citenamefont {Wang},\ and\ \citenamefont
  {Jiang}}]{Peng2021}%
  \BibitemOpen
  \bibfield  {author} {\bibinfo {author} {\bibfnamefont {C.}~\bibnamefont
  {Peng}}, \bibinfo {author} {\bibfnamefont {Y.-F.}\ \bibnamefont {Jiang}},
  \bibinfo {author} {\bibfnamefont {Y.}~\bibnamefont {Wang}}, \ and\ \bibinfo
  {author} {\bibfnamefont {H.-C.}\ \bibnamefont {Jiang}},\ }\href {\doibase
  10.1088/1367-2630/ac3a83} {\bibfield  {journal} {\bibinfo  {journal} {New
  Journal of Physics}\ }\textbf {\bibinfo {volume} {23}},\ \bibinfo {pages}
  {123004} (\bibinfo {year} {2021})}\BibitemShut {NoStop}%
\bibitem [{\citenamefont {Zhu}\ \emph {et~al.}(2022)\citenamefont {Zhu},
  \citenamefont {Sheng},\ and\ \citenamefont {Vishwanath}}]{zhu2022}%
  \BibitemOpen
  \bibfield  {author} {\bibinfo {author} {\bibfnamefont {Z.}~\bibnamefont
  {Zhu}}, \bibinfo {author} {\bibfnamefont {D.~N.}\ \bibnamefont {Sheng}}, \
  and\ \bibinfo {author} {\bibfnamefont {A.}~\bibnamefont {Vishwanath}},\
  }\href {\doibase 10.1103/PhysRevB.105.205110} {\bibfield  {journal} {\bibinfo
   {journal} {Phys. Rev. B}\ }\textbf {\bibinfo {volume} {105}},\ \bibinfo
  {pages} {205110} (\bibinfo {year} {2022})}\BibitemShut {NoStop}%
\bibitem [{\citenamefont {Huang}\ \emph {et~al.}(2022)\citenamefont {Huang},
  \citenamefont {Gong},\ and\ \citenamefont {Sheng}}]{Huang2022}%
  \BibitemOpen
  \bibfield  {author} {\bibinfo {author} {\bibfnamefont {Y.}~\bibnamefont
  {Huang}}, \bibinfo {author} {\bibfnamefont {S.-S.}\ \bibnamefont {Gong}}, \
  and\ \bibinfo {author} {\bibfnamefont {D.~N.}\ \bibnamefont {Sheng}},\ }\href
  {https://arxiv.org/abs/2209.00833} {\bibfield  {journal} {\bibinfo  {journal}
  {arxiv:2209.00833}\ } (\bibinfo {year} {2022})}\BibitemShut {NoStop}%
\bibitem [{\citenamefont {Jiang}\ and\ \citenamefont
  {Kivelson}(2022)}]{Jiang2022pnas}%
  \BibitemOpen
  \bibfield  {author} {\bibinfo {author} {\bibfnamefont {H.-C.}\ \bibnamefont
  {Jiang}}\ and\ \bibinfo {author} {\bibfnamefont {S.~A.}\ \bibnamefont
  {Kivelson}},\ }\href {\doibase 10.1073/pnas.2109406119} {\bibfield  {journal}
  {\bibinfo  {journal} {Proceedings of the National Academy of Sciences}\
  }\textbf {\bibinfo {volume} {119}},\ \bibinfo {pages} {e2109406119} (\bibinfo
  {year} {2022})}\BibitemShut {NoStop}%
\bibitem [{\citenamefont {Peng}\ \emph {et~al.}(2022)\citenamefont {Peng},
  \citenamefont {Wang}, \citenamefont {Wen}, \citenamefont {Lee}, \citenamefont
  {Devereaux},\ and\ \citenamefont {Jiang}}]{Peng2022}%
  \BibitemOpen
  \bibfield  {author} {\bibinfo {author} {\bibfnamefont {C.}~\bibnamefont
  {Peng}}, \bibinfo {author} {\bibfnamefont {Y.}~\bibnamefont {Wang}}, \bibinfo
  {author} {\bibfnamefont {J.}~\bibnamefont {Wen}}, \bibinfo {author}
  {\bibfnamefont {Y.}~\bibnamefont {Lee}}, \bibinfo {author} {\bibfnamefont
  {T.}~\bibnamefont {Devereaux}}, \ and\ \bibinfo {author} {\bibfnamefont
  {H.-C.}\ \bibnamefont {Jiang}},\ }\href {https://arxiv.org/abs/2206.03486}
  {\bibfield  {journal} {\bibinfo  {journal} {arXiv:2206.03486}\ } (\bibinfo
  {year} {2022})}\BibitemShut {NoStop}%
\end{thebibliography}
%

\end{document}